# Unidirectional imaging with partially coherent light

Guangdong Ma[1,2,3,4,†], Che-Yung Shen[1,2,3,†], Jingxi Li[1,2,3], Luzhe Huang[1,2,3], Çağatay Işıl[1,2,3], Fazil Onuralp Ardic[1], Xilin Yang[1,2,3], Yuhang Li[1,2,3], Yuntian Wang[1,2,3], Md Sadman Sakib Rahman[1,2,3], Aydogan Ozcan[*,1,2,3,5]

[1]Electrical and Computer Engineering Department, University of California, Los Angeles, CA, 90095, USA.

[2]Bioengineering Department, University of California, Los Angeles, 90095, USA.

[3]California NanoSystems Institute (CNSI), University of California, Los Angeles, CA, 90095, USA.

[4]School of Physics, Xi'an Jiaotong University, Xi'an, Shaanxi, 710049, China.

[5]Department of Surgery, University of California, Los Angeles, CA, 90095, USA.

†These authors contributed equally

*Correspondence: Aydogan Ozcan, ozcan@ucla.edu

**Abstract**

Unidirectional imagers form images of input objects only in one direction, e.g., from field-of-view (FOV) A to FOV B, while blocking the image formation in the reverse direction, from FOV B to FOV A. Here, we report unidirectional imaging under spatially partially coherent light and demonstrate high-quality imaging only in the forward direction (A→B) with high power efficiency while distorting the image formation in the backward direction (B→A) along with low power efficiency. Our reciprocal design features a set of spatially engineered linear diffractive layers that are statistically optimized for partially coherent illumination with a given phase correlation length. Our analyses reveal that when illuminated by a partially coherent beam with a correlation length of $\geq \sim 1.5\lambda$, where $\lambda$ is the wavelength of light, diffractive unidirectional imagers achieve robust performance, exhibiting asymmetric imaging performance between the forward and backward directions – as desired. A partially coherent unidirectional imager designed with a smaller correlation length of $< 1.5\lambda$ still supports unidirectional image transmission, but with a reduced figure of merit. These partially coherent diffractive unidirectional imagers are compact (axially spanning $<75\lambda$), polarization-independent, and compatible with various types of illumination sources, making them well-suited for applications in asymmetric visual information processing and communication.

## Introduction

Controlling and engineering the properties and behavior of light as a function of the wave propagation direction has been crucial for various advancements in optical sensing and imaging[1,2], including e.g., unidirectional wave transmission systems. Common strategies for unidirectional transmission include employing e.g., temporal modulation, magneto-optical effect, nonlinear materials, or multi-layer spatial modulation of light [3–12]. For example, nonlinear optical materials with intensity-dependent permittivity can be used to create nonreciprocal devices[3–7]. As another example, the engineering of structural asymmetry introduced by multi-layer, lossy linear diffractive systems, despite being reciprocal, can also be used to create asymmetric wave transmission under spatially coherent illumination[8–12]. These strategies have been used for the unidirectional transmission of waves, ensuring high-fidelity delivery of forward signals while featuring losses and distortions for backward signals. Nevertheless, these existing methods typically require high-power beams or have relied on spatially coherent illumination.

Partially coherent light, in general, helps mitigate image degradation due to speckle noise, minimizes crosstalk among imaged objects, and is less susceptible to misalignments or defects in the optical system. Some of these benefits make partially coherent illumination particularly attractive for mobile microscopy[13–15], quantitative phase imaging[16], virtual and augmented reality displays[17], and light beam shaping[18], among other applications[19].

Here, we present unidirectional diffractive imagers that operate under spatially partially coherent illumination, featuring high image quality and power efficiency in the forward direction (A→B), while distorting the image formation in the backward direction (B→A) along with reduced power efficiency. Each partially coherent unidirectional imager represents a reciprocal and lossy linear optical device, designed through a set of spatially engineered diffractive layers that are jointly optimized using deep learning[8,20–22]. Our findings indicate that the degree of the spatial coherence of the illumination, statistically quantified with the phase correlation length ($C_\phi$), significantly impacts the performance of the unidirectional imager. Specifically, unidirectional diffractive imagers designed with a training correlation length ($C_\phi^{\mathrm{train}}$) greater than $\sim 1.5\lambda$ exhibit very good unidirectional imaging behavior, accurately reproducing the input images with high structural fidelity and diffraction efficiency in the forward direction (A → B), while suppressing image formation in the reverse path (B → A). Diffractive unidirectional imagers designed for a smaller correlation length of $< 1.5\lambda$ still maintain asymmetric image formation, however, with a reduced figure of merit (FOM).

We validated the unidirectional imaging performance of diffractive visual processors using different levels of spatial coherence, even though they were trained with a specific phase correlation length ($C_{\phi}^{\text{train}}$), demonstrating the resilience of partially coherent unidirectional imagers to unknown changes in the spatial coherence properties of the illumination beam. We also demonstrated the internal and external generalization of the unidirectional imager designs across various image datasets, further highlighting their resilience to unknown data distribution shifts. With the unique advantages of being compact (with an axial length of <75λ), polarization-insensitive, and compatible with different types of partially coherent sources, including light-emitting-diodes (LEDs), the presented unidirectional imager design offers new capabilities for asymmetric visual information processing.

## Results

### Unidirectional imager design under spatially partially coherent illumination

Figure 1 (a) illustrates the concept of a unidirectional imager under a spatially partially coherent, monochromatic illumination at a wavelength of $\lambda$. The processor is designed to implement a unidirectional imaging task: high structural similarity and high diffraction efficiency for the forward operation A → B (blue line in Figure 1 a) together with a distorted image and reduced diffraction efficiency at the backward operation B → A (brown line in Figure 1 a). We used four diffractive layers, axially spaced by $d$ in this design. Each diffractive layer contains $200 \times 200$ diffractive features used to modulate the phase of the transmitted optical field, with each feature having a lateral size of $0.53\lambda$ and a trainable phase value covering $[0, 2\pi]$. The transmission functions of these diffractive features are optimized using a training dataset composed of MNIST handwritten digits (see the Methods section). An engineered loss function guides the optimization of the unidirectional visual processor toward achieving two primary goals: (1) for A → B, it aims to minimize the structural differences between the output images and the ground truth images using the normalized mean square error (NMSE) and Pearson Correlation Coefficient (PCC), while concurrently maximizing the forward diffraction efficiency; (2) for B → A, the loss function maximizes the differences between the backward output images and the ground truth images and simultaneously minimizes the backward diffraction efficiency (refer to the Methods for details).

For spatially partially coherent monochromatic illumination, we used the phase correlation length, $C_{\phi}$, to quantify the degree of the spatial coherence of the source[23,24]. Figure 1 (b) illustrates some examples of the random phase profiles of partially coherent illumination with different correlation lengths, varying from $\sim 0.5\lambda$ to ~$3\lambda$; also see the Methods section and Supplementary Figure S1 for details. During the training of a unidirectional diffractive imager, for each input object, we use $\mathrm{N}_{\phi}^{\mathrm{train}}$ different random phase profiles that follow a given correlation length $\mathrm{C}_{\phi}$ at the input plane; the time-averaged intensity of the resulting complex output fields for these $\mathrm{N}_{\phi}^{\mathrm{train}}$ different phase profiles is then used to optimize the unidirectional imager performance based on our training loss function. Details about the optical model of a diffractive unidirectional imager, the training strategy, and loss functions can be found in the Methods section.

Figure 2 (a) illustrates the layout of the partially coherent unidirectional imager design with a compact axial length of $\sim 75\lambda$, which was optimized using $\mathrm{N}_{\phi}^{\mathrm{train}} = 16$ and $\mathrm{C}_{\phi}^{\mathrm{train}} = 2.5\lambda$. This deep learning-optimized diffractive design is composed of four spatially engineered diffractive layers that are displayed in Figure 2 (b). We blindly tested this partially coherent unidirectional imager using 10,000 handwritten digits never seen during the training process; for each unknown input test object, 2,048 random phase profiles, each with a phase correlation length of 2.5λ, were used to obtain the time-averaged intensity at the output plane (i.e., $\mathrm{N}_{\phi}^{\mathrm{test}} = 2048$ – which remained the same throughout our manuscript). We refer to this testing scheme as "internal generalization" from the perspective of the spatial coherence properties of the illumination light since we maintained the same statistical phase correlation length in the testing stage as used in the training, i.e., $\mathrm{C}_{\phi}^{\mathrm{train}} = \mathrm{C}_{\phi}^{\mathrm{test}} = 2.5\lambda$. Some of these blind testing results are illustrated in Figure 2 (c). The first row in Figure 2 (c) shows the input amplitude objects used for both the forward and backward directions. The following two rows in Figure 2 (c) show the forward output images and the backward output images. To better illustrate the details of the backward output images, the last row in Figure 2 (c) further displays higher contrast images of the backward direction B → A, covering a lower intensity range. These visual comparisons clearly demonstrate the success of the partially coherent diffractive unidirectional imager design, reproducing the input images in the forward direction while blocking them in the backward direction – as desired. We also quantified the performance of this unidirectional visual processor using various metrics, including PCC, diffraction efficiency, and PSNR of the forward and backward directions, as shown in Figures 2 (d) – 2 (f). These metrics are calculated across 10,000 MNIST test objects, never used in training. The PCC values for the forward and backward are 0.9541 $\pm$ 0.0239, 0.1464 $\pm$ 0.1005, respectively. Furthermore, the forward diffraction efficiency, 85.59%$\pm$0.05%, is about four-fold higher than the backward diffraction efficiency, 22.41%$\pm$0.03%, as shown in Figure 2 (e). A similar desired performance is also observed between the forward PSNR and backward PSNR (18.46$\pm$1.84 and 8.79$\pm$2.01, respectively), as shown in Figure 2 (f).

**Impact of $N_{\phi}^{train}$ on the performance of partially coherent unidirectional diffractive imagers**

To explore the impact of $N_{\phi}^{train}$ on the performance of partially coherent unidirectional imagers, we trained seven diffractive processors with different $N_{\phi}^{train}$ values ranging from 1 to 64; see Figure 3. All these diffractive models were trained and tested using the same partially coherent illumination with $C_{\phi}^{train} = C_{\phi}^{test} = 2.5\lambda$ and in our blind testing, we used $N_{\phi}^{test} = 2048$. We observed that the diffractive processors trained with a larger $N_{\phi}^{train}$ exhibited improved asymmetric imaging performance between the forward and backward directions, as quantified in Figures 3(a-c). To better quantify the unidirectional imaging capability of a partially coherent diffractive processor, we defined an FOM by calculating the sum of the diffraction efficiency ratio and the image PSNR ratio between the forward and backward imaging directions (see the Methods section). The resulting FOM is reported as a function of $N_{\phi}^{train}$ in Figure 3 (d), which reveals that increasing $N_{\phi}^{train}$ improves the FOM of the unidirectional imager up to $N_{\phi}^{train} \sim 16$ and beyond 16 the performance differences among different designs diminish. Figure 3 (e) further presents the blind testing results for different $N_{\phi}^{train}$ values. The first three rows of Figure 3 (e) display the input amplitude objects (used for both the forward and backward directions), the forward outputs, and the backward outputs, respectively. The last row in Figure 3 (e) also displays the backward outputs with increased contrast. As desired, all the backward output images of these diffractive models appear as noise with poor diffraction efficiency. Based on these performance analyses, we conclude that $N_{\phi}^{train} = 16$ is sufficient to design/train a partially coherent diffractive unidirectional imager, and therefore, we adopted $N_{\phi}^{train} = 16$ in subsequent diffractive models to speed up the training process.

Furthermore, Supplementary Figure S2 illustrates the numerical performance comparison of a diffractive model trained with $N_{\phi}^{train} = 16$ and tested under different $N_{\phi}^{test}$ values, revealing similar performance results for $N_{\phi}^{test}$ ranging from 16 to 2048. In the rest of our blind testing analysis, we used $N_{\phi}^{test} = 2048$ since the testing time is negligible compared to the training process.

**Impact of $C_{\phi}^{train}$ on the performance of partially coherent unidirectional diffractive imagers**

In previous sub-sections, we analyzed the performance of partially coherent unidirectional diffractive imager designs with $C_{\phi}^{train} = C_{\phi}^{test} = 2.5\lambda$. To understand the impact of the $C_{\phi}^{train}$ on the performance of unidirectional imaging, we trained six additional diffractive visual processors under partially coherent illumination with different correlation lengths ranging from $\sim 0.5\lambda$ to ~$3\lambda$. Each diffractive visual processor was tested with the same phase correlation length as used in the training, i.e., $C_{\phi}^{train} = C_{\phi}^{test}$; see Figure

4. To evaluate the performances of these diffractive visual processors in both the forward and backward directions, we calculated various metrics, such as PCC, diffraction efficiency, PSNR, and FOM, using 10,000 MNIST test images; see Figures 4 (a-d). A visualization of the blind testing results is also displayed in Figure 4 (e), where the first three rows depict the input amplitude objects, the forward outputs, and the backward outputs, respectively. For better comparison, the last row in Figure 4 (e) also shows the backward output images with increased contrast. Note that we use a reduced intensity range for diffractive models with smaller phase correlation lengths due to their poor diffraction efficiency, as indicated by the red box in Figure 4 (e).

According to the FOM values presented in Figure 4 (d) and the visualizations in Figure 4 (e), partially coherent diffractive processors exhibit a very good unidirectional imaging performance (FOM ≥ 4) when trained with a larger correlation length of $C_{\phi}^{\text{train}} \geq 1.5\lambda$, enabling unidirectional imaging with a large diffraction efficiency in the forward direction while suppressing image formation with a significantly reduced diffraction efficiency in the opposite direction. However, for diffractive processors trained with $C_{\phi}^{\text{train}} < 1.5\lambda$, the unidirectional imaging performance appears to diminish. Specifically, both the forward and backward output diffraction efficiencies of these diffractive models trained with $C_{\phi}^{\text{train}} < 1.5\lambda$ fall below $1\%$, accompanied by a reduced FOM of ~1-2 (Fig. 4 d). When illuminated with partially coherent light with shorter correlation lengths, images of the input patterns can be observed in both the forward and backward outputs.

To further explore the design characteristics of diffractive unidirectional imagers trained under different correlation lengths ranging from $0.5\lambda$ to $3\lambda$, Figure 5 illustrates the phase profiles of the optimized diffractive layers of each design. We observe that the central parts of the diffractive layers trained using $C_{\phi}^{\text{train}} = 0.5\lambda$ and $C_{\phi}^{\text{train}} = 1.0\lambda$ are relatively flat, which indicates poor imaging performance since these central parts are crucial for image formation. In contrast, the diffractive layers trained with larger correlation lengths, $C_{\phi}^{\text{train}} \geq 1.5\lambda$, exhibit a completely different topology in each diffractive layer, indicating better learning/convergence and more effective utilization of the independent degrees of freedom at each diffractive layer, which is at the heart of the better unidirectional imaging performance achieved for these diffractive models trained with $C_{\phi}^{\text{train}} \geq 1.5\lambda$, as also illustrated in Fig. 4.

So far, we used the same level of partial spatial coherence during both the training and blind testing stages, i.e., $C_{\phi}^{\text{train}} = C_{\phi}^{\text{test}}$. Next, we explored the generalization performance of a unidirectional diffractive imager under different levels of partial coherence during the blind testing stage; see Figure 6. Each line in Figure 6 depicts the performance of a unidirectional imager design (trained using a specific $C_{\phi}^{train}$) with respect to varying $C_{\phi}^{test}$ values. These findings support our previous observations, revealing that the unidirectional diffractive imagers trained with $C_{\phi}^{train} \geq 1.5\lambda$ perform well, and these diffractive designs do not necessarily overfit to a specific $C_{\phi}^{train}$ value, exhibiting improved FOM as long as $C_{\phi}^{test} \geq 1.5\lambda$; also see Supplementary Figures S3 to S5 for some blind testing examples further supporting these conclusions.

In Figure 6 and Supplementary Figures S4-S5, we also observe that none of these diffractive designs achieve a decent unidirectional imaging FOM when tested with $C_{\phi}^{test} = 0.5\lambda$ and $C_{\phi}^{test} = 1.0\lambda$, i.e., when the phase correlation length of the illumination beam approaches the diffraction limit of light in air. This poor performance of the unidirectional imager designs reported in Fig. 6 with $C_{\phi}^{test} < C_{\phi}^{train} \geq 1.5\lambda$ is due to the fact the diffractive layers shown in Fig. 5 for $C_{\phi}^{train} \geq 1.5\lambda$ overfitted to the relatively larger spatial coherence diameter of the illumination, failing external generalization and unidirectional imaging for $C_{\phi}^{test} = 0.5\lambda$ and $C_{\phi}^{test} = 1.0\lambda$. However, the failure of the diffractive designs with $C_{\phi}^{test} = C_{\phi}^{train} = 0.5\lambda$ and $C_{\phi}^{test} = C_{\phi}^{train} = 1\lambda$ can be attributed to the lower spatial resolution and relative sparsity of our training dataset, failing to cover phase correlation lengths closer to the diffraction limit of light. To better shed light on this, we trained a new diffractive unidirectional imager ($C_{\phi}^{train} = C_{\phi}^{test} = 0.5\lambda$) with a higher resolution training image dataset featuring random intensity patterns; see Supplementary Fig. S6 and the Methods section for details. Supplementary Figure S6(e) reveals that the diffractive layers of this new unidirectional imager design better utilize the independent degrees of freedom at each layer, and avoid relatively smooth large regions at the center of each layer, which indicates a better optimization of the unidirectional imager. This improved performance is also evident in the comparisons provided in Supplementary Figs. S6(a-d) and Fig. 4e, where the FOM of unidirectional imaging increased to 3.32, an improvement of >2.2-fold compared to our earlier design with $C_{\phi}^{train} = C_{\phi}^{test} = 0.5\lambda$. Supplementary Figure S6 also reports unidirectional imaging metrics of this design, i.e., PCC, diffraction efficiency, and PSNR, which are calculated using 10,000 random test images, further supporting the improved performance of this new spatially incoherent design. These analyses underscore the crucial role of the training image dataset in the performance of a diffractive unidirectional imager, especially for a phase correlation length of $< \lambda$.

**External generalization of partially coherent unidirectional diffractive imager designs to new image datasets**

Next, we showcase the external generalization capabilities of partially coherent diffractive unidirectional imager designs to other datasets that were never used before. For this analysis, we used the EMNIST dataset[25] containing images of handwritten English letters, and a customized image dataset featuring various types of gratings. Both of these were never used during the training, which only used handwritten digits. These external generalization test results shown in Figure 7 are obtained using the unidirectional imager design with $C_{\phi}^{\text{train}} = C_{\phi}^{\text{test}} = 2.5\lambda$. The first three rows in Figures 7 (a) and 7 (b) depict the input amplitude objects, the forward outputs, and the backward outputs, respectively, all using the same intensity range. The last row shows the backward output images with increased image contrast, clearly confirming the unidirectional imaging capability and the successful external generalization of the diffractive design to unseen input images from different datasets.

We also quantified the spatial resolution of this partially coherent unidirectional imager ($C_{\phi}^{\text{train}} = C_{\phi}^{\text{test}} = 2.5\lambda$) using various resolution test targets with different linewidths that were previously unseen (see Figure 8). The minimum resolvable linewidth (period) by this unidirectional imager design was found to be ~$1\lambda$ (~$2\lambda$), as indicated by the results in Figure 8. These results further validate the successful external generalization of the unidirectional imager design for general-purpose imaging operation exclusively in the forward direction, despite being trained solely using the MNIST image dataset.

**Discussion**

In this work, we introduced a unidirectional diffractive imager that works under spatially partially coherent light, designed to image in one direction with high power efficiency while blocking/suppressing the image formation in the opposite direction with reduced diffraction efficiency. The presented unidirectional imager comprises phase-only diffractive layers optimized by deep learning, and it axially spans only <75λ, making it very compact. Our analyses revealed that when the phase correlation length of the illumination source exceeds ~$1.5\lambda$, the partially coherent unidirectional processor designs exhibit a very good unidirectional imaging performance with an FOM of >4, enabling high diffraction efficiency imaging in the forward direction while inhibiting the image formation in the opposite direction with reduced diffraction efficiency. However, diffractive processors trained with $C_{\phi}^{\text{train}} < 1.5\lambda$ show diminished asymmetric transmission, with both the forward and backward output diffraction efficiencies falling below $1\%$ along with an FOM of ~1-2. As a mitigation strategy, we demonstrated that this performance limitation can be addressed using a higher resolution training image dataset, which improved the unidirectional imaging

FOM to >3.3 even for $C_{\phi}^{train} = C_{\phi}^{test} = 0.5\lambda$. Furthermore, we successfully demonstrated both the internal and external generalizability of our unidirectional imager designs across different image datasets.

Being a reciprocal optical design, the asymmetric information transmission that is achieved by diffractive linear optical processors is based on the task-specific engineering of the forward and backward point spread functions (PSFs) that are spatially varying[26–29]; this conclusion is true for spatially coherent, spatially incoherent or partially coherent diffractive optical processors. In general, a diffractive optical processor can be trained, through image data, to approximate any arbitrary set of spatially varying PSFs between an input and output FOV. For example, under spatially coherent illumination, with sufficient degrees of freedom within the diffractive layers, a diffractive processor can approximate any arbitrary PSF set, $h(m, n; m', n')$, where $h$ is the spatially coherent complex-valued PSF, and $(m, n)$ and $(m', n')$ define the coordinates of the output and input FOVs, respectively. Similarly, under spatially incoherent illumination, any arbitrary spatially varying intensity impulse response, $H(m, n; m', n') = |h(m, n; m', n')|^2$, can be approximated through data-driven learning[30]. The loss function and the training image data are important for the accuracy, spatial resolution and generalization behavior of these linear transformations. For the context of unidirectional imaging using lossy diffractive linear processors, however, the goal is to engineer the forward and backward spatially varying PSFs and make them asymmetric, suppressing the image formation in the backward direction while maintaining decent images in the forward direction. There is no unique solution for this task since infinitely many combinations of $h_F(m, n; m', n')$ and $h_B(m', n'; m, n)$ can be devised to achieve a desired unidirectional FOM value, where $h_F$ and $h_B$ refer to the spatially coherent PSF sets for the forward and backward directions, respectively. It should be emphasized that spatially incoherent and partially coherent diffractive unidirectional imagers can all be modeled through the behavior of $h_F$ and $h_B$ under statistically varying illumination phase patterns (defined by $C_{\phi}$); see the Methods section.

In addition to the degree of coherence of the incident light and the spatial features of the training image dataset, the performance of a unidirectional imager design is also influenced by several system properties, including the number of diffractive layers, the number of trainable features in each diffractive layer, the wavelength, the bandwidth of the source, the axial distance between successive layers, and the pixel pitch. Increasing the number of trainable parameters, such as the number of layers and/or the number of diffractive features, enhances the overall performance of the system, albeit at the cost of longer training and fabrication/assembly time[26,27]. Furthermore, by physically adjusting the structure of the diffractive layers, the presented design can be specifically tailored to perform unidirectional imaging with a desired magnification or demagnification factor[12].

Finally, the unidirectional imagers introduced in this work are highly compact, axially spanning <75λ, and they exhibit significant versatility that can be adapted to various parts of the electromagnetic spectrum; by scaling the resulting diffractive features of each transmissive layer proportional to the illumination wavelength, the same design can operate at different parts of the spectrum, including the visible and infrared wavelengths – *without* the need to re-design the diffractive layers of the unidirectional imager. This adaptability is poised to facilitate various novel applications in diverse fields, such as asymmetric visual information processing and communication, potentially enhancing privacy protection and mitigating multipath interference within optical communication systems, among others.

## Methods

### Forward model of a diffractive visual processor under spatially coherent illumination

The propagation of a coherent field from the $lth$ to the $(l+1)th$ diffractive plane is calculated using the angular spectrum method [31]:

$$u^{l+1}(x,y) = \mathbb{P}_d u^l(x,y) = \mathcal{F}^{-1}\{\mathcal{F}\{u^l(x,y)\}H(f_x,f_y,d)\} \quad (1),$$

where $\mathbb{P}_d$ denotes the free-space propagation operation, and $d$ represents the axial distance between two successive planes. $\mathcal{F}\{\cdot\}$ and $\mathcal{F}^{-1}\{\cdot\}$ represent the 2D Fourier transform and the inverse Fourier transform operations, respectively. The transfer function $H(f_x,f_y,d)$ is defined as:

$$H(f_x,f_y,d) = \begin{cases} \exp(\mathrm{j}2\pi d\sqrt{1/\lambda^2 - f_x^2 - f_y^2}), & 1/\lambda^2 - f_x^2 - f_y^2 > 0 \\ 0, & 1/\lambda^2 - f_x^2 - f_y^2 \le 0 \end{cases} \quad (2),$$

where $\mathrm{j} = \sqrt{-1}$ and $k = \frac{2\pi}{\lambda}$. $f_x$ and $f_y$ denote the spatial frequencies along the $x$ and y directions. The $l$th diffractive layer modulates the phase of the transmitted optical field with a transmission function, $t^l(x,y)$:

$$t^l(x,y) = \exp\left(\mathrm{j}\phi^l(x,y)\right) \quad (3),$$

where $\phi^l(x,y)$ denotes the learnable phase profile of the diffractive features located at the $l$th diffractive layer. The output intensity $\hat{O}_i(x,y)$ of a K-layer diffractive visual processor can be written as:

$$\hat{O}_i(x,y) = \left|\mathbb{P}_d\left(\left(\prod_{l=1}^{K} t^l(x,y)\,\mathbb{P}_d\right)u_i^0(x,y)\right)\right|^2 \tag{4}$$

Here, $u_i^0$ is the complex field at the input FOV:

$$u_i^0(x,y) = \sqrt{I}\exp\left(\mathrm{j}\varphi_i(x,y)\right) \tag{5}$$

where $\varphi_i(x,y)$ represents the phase profile of the incident optical field and $I$ refers to the intensity profile.

**Forward model of a diffractive visual processor under spatially partially coherent illumination**

To model the propagation of a partially coherent field, we define the phase profile $\varphi_i(x,y)$ of the incident field as [23,32,33]:

$$\varphi_i(x,y) = \mathrm{mod}\left(\frac{2\pi}{\lambda}W(x,y) * G(x,y), 2\pi\right) \tag{6}$$

where $\lambda$ is the wavelength of the illumination light. '$*$' refers to the 2D convolution operation. $W(x,y)$ follows a normal distribution with a mean value of $\mu$ and a standard deviation of $\sigma_0$, i.e., $W(x,y) \sim N(\mu,\sigma_0)$. $G(x,y)$ represents a zero-mean Gaussian smoothing kernel with a standard deviation of $\sigma$, defined by $\exp(-(x^2+y^2)/2\sigma^2)$. We numerically tailored these phase profiles to the desired correlation length by adjusting the standard deviation, $\sigma$, of the Gaussian smoothing kernel. The phase correlation length, $C_\phi$, of a partially coherent field was approximated using the autocorrelation function, $R_\phi(x,y)$, of the phase profile, $\varphi_i(x,y)$[23,34,35]:

$$R_\phi(x,y) = \mathcal{F}^{-1}\{|\mathcal{F}\{\varphi_i(x,y)\}|^2\} = \exp\left(-\pi(x^2+y^2)/C_\phi\right) \tag{7}$$

Using Eq. (7), for a given combination of $\mu$, $\sigma_0$ and $\sigma$ values, we numerically approximated $C_\phi$ based on 2048 randomly selected phase profiles, $\varphi_i(x,y)$. In this study, we used $C_\phi$ values ranging from ~$0.5\lambda$ to ~$3.0\lambda$ with increments of ~$0.5\lambda$, corresponding to $\sigma$ values of ~$1.9\lambda$, ~$2.4\lambda$, ~$3.9\lambda$, ~$4.5\lambda$, and ~$5.0\lambda$, respectively. Supplementary Figure S1 provides more information on the generation of these phase profiles with the above-described parameters.

The time-averaged intensity $\hat{O}(x,y)$ of the diffractive visual processor under spatially partially coherent illumination is calculated as follows:

$$\hat{O}(x,y) = \langle\hat{O}_i(x,y)\rangle = \lim_{N_\phi\to\infty}\frac{1}{N_\phi}\sum_{i=1}^{N_\phi}\hat{O}_i(x,y) \tag{8}$$

Due to limited computing resources, we used $N_\phi^{\mathrm{test}} = 2048$ in the blind testing stage of each trained diffractive model.

### Training details

The unidirectional diffractive imagers reported in this work are designed for spatially partially coherent illumination at a wavelength of $\lambda = 0.75$ mm. The FOV A and FOV B (Figure 1) share the same physical size of $11.2\text{mm} \times 11.2\text{mm}$ (i.e., $\sim 15\lambda \times 15\lambda$), discretized into $28 \times 28$-pixels. Each diffractive layer contains $200 \times 200$ trainable diffractive features, modulating only the phase of the transmitted field. The axial distance between any two successive diffractive planes of a unidirectional imager is set to $11$ mm, i.e., $\mathrm{d} \approx 14.67\lambda$, corresponding to a numerical aperture of 0.96 within the diffractive system.

All the unidirectional diffractive imagers were optimized using a training dataset composed of 55,000 MNIST handwritten digits, except for one diffractive processor illustrated in Supplementary Figure S6. To enhance their generalization capabilities, we randomly applied dilation or erosion operations to the original MNIST images using OpenCV's built-in functions, 'cv2.dilate' or 'cv2.erode', respectively. After data augmentation, the dilated, eroded, and original MNIST images were combined into a mixed dataset. This dataset was then divided into training, validation, and testing sets, each containing 55,000, 5,000, and 10,000 images, respectively, with no overlap. Note that the unidirectional visual processor depicted in Supplementary Figure S6 was trained using a higher resolution image dataset consisting random intensity patterns within an intensity range of [0, 1]. The total number of images in this random image dataset is equivalent to that of the MNIST dataset.

All the diffractive visual processors are trained using the default optimizer (optax) in JAX, with a learning rate of 0.001 and a batch size of 32, over 50 epochs. All the models were trained and tested on JAX (version 0.4.1), utilizing a GeForce RTX 4090 graphical processing unit (GPU) from NVIDIA Inc. Training a partially coherent diffractive unidirectional imager with four diffractive layers typically takes ~5 hours.

### Performance evaluation metrics

To evaluate the performance of each unidirectional diffractive imager, four different metrics are used:

1) The PCC value between the forward (or backward) output of the unidirectional diffractive processor and the ground truth image.

2) The forward (or backward) diffraction efficiency.

3) PSNR between the forward (or backward) output image and the target intensity, $I$, defined as:

$$\mathrm{PSNR}(\hat{O}_{for}, I) = 10\log10\left(\frac{255^2}{\mathrm{NMSE}(\hat{O}_{for}, I)}\right) .$$

4) The FOM of the unidirectional diffractive imager:

$$\mathrm{FOM}(\hat{O}_{for}, \hat{O}_{back}, I) = \frac{\eta(\hat{O}_{for}, I)}{\eta(\hat{O}_{back}, I)} + \frac{\mathrm{PSNR}(\hat{O}_{for}, I)}{\mathrm{PSNR}(\hat{O}_{back}, I)}.$$

When calculating the backward metrics, such as $\mathrm{PCC}(\hat{O}_{back}, I)$, $\eta(\hat{O}_{back}, I)$, and $\mathrm{PSNR}(\hat{O}_{back}, I)$, we simply replaced $\hat{O}_{for}$ with $\hat{O}_{back}$ while keeping the other parameters unchanged.

**Supporting Information**: This file contains:

- Supplementary Figures S1-S6.
- Training loss function

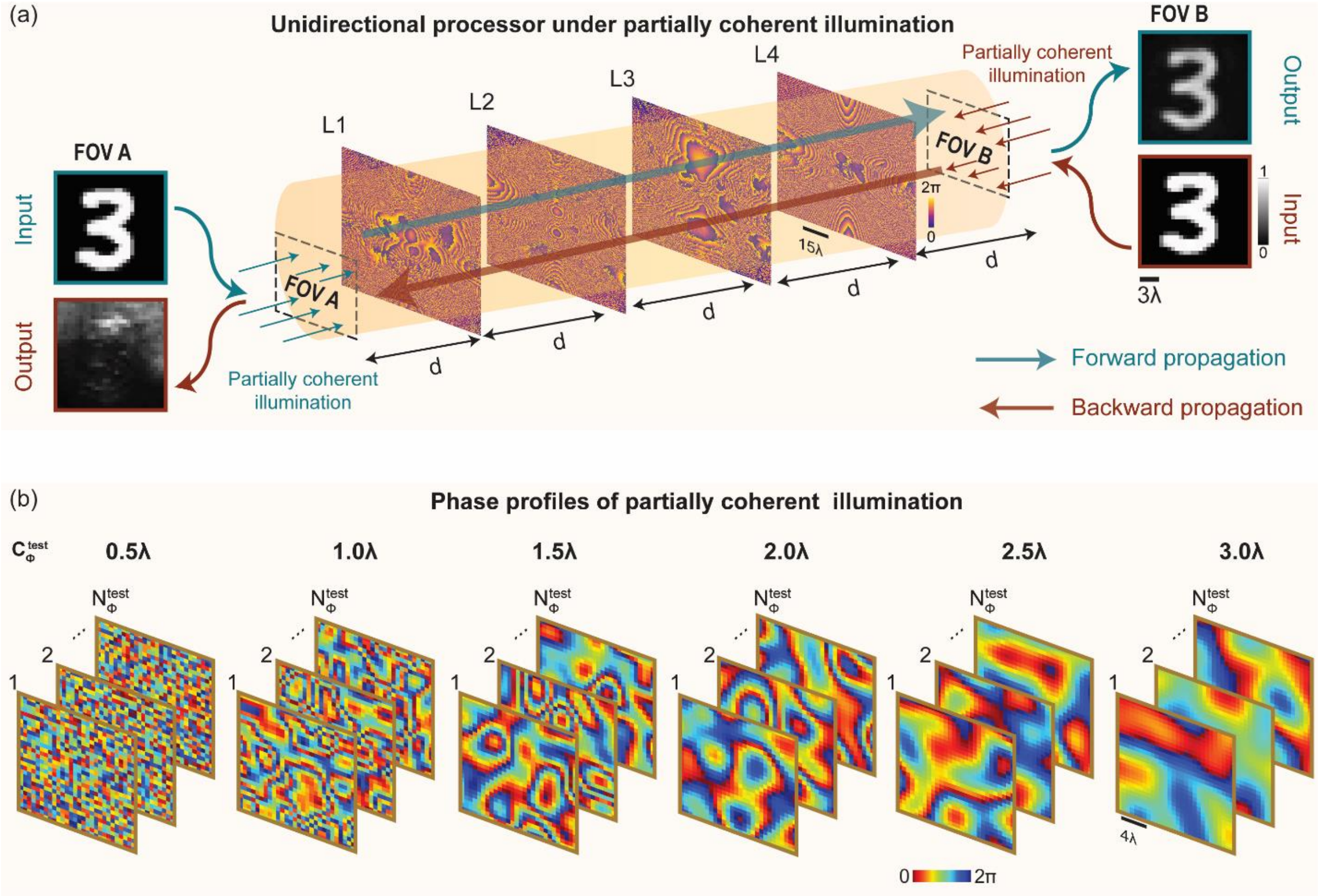

**Figure 1. Concept of a unidirectional diffractive imager with partially coherent illumination.** (a) Schematic of unidirectional imager under partially coherent, monochromatic illumination with a wavelength of $\lambda$. The unidirectional diffractive processor reproduces the input object's image in the forward propagation direction (blue line from FOV A to FOV B), while suppressing the image formation in the backward propagation direction (brown line from FOV B to FOV A). This design includes four phase-only diffractive layers axially spaced by $d$, each containing 200 × 200 diffractive features that modulate the phase of the transmitted optical field. (b) Six sets of random phase profiles of partially coherent illumination, each containing $\mathrm{N}_{\Phi}^{\mathrm{test}}$ phase profiles. Each set corresponds to one specific correlation length, $\mathrm{C}_{\Phi}^{\mathrm{test}}$ , varying from ~0.5$\lambda$ to ~3$\lambda$.

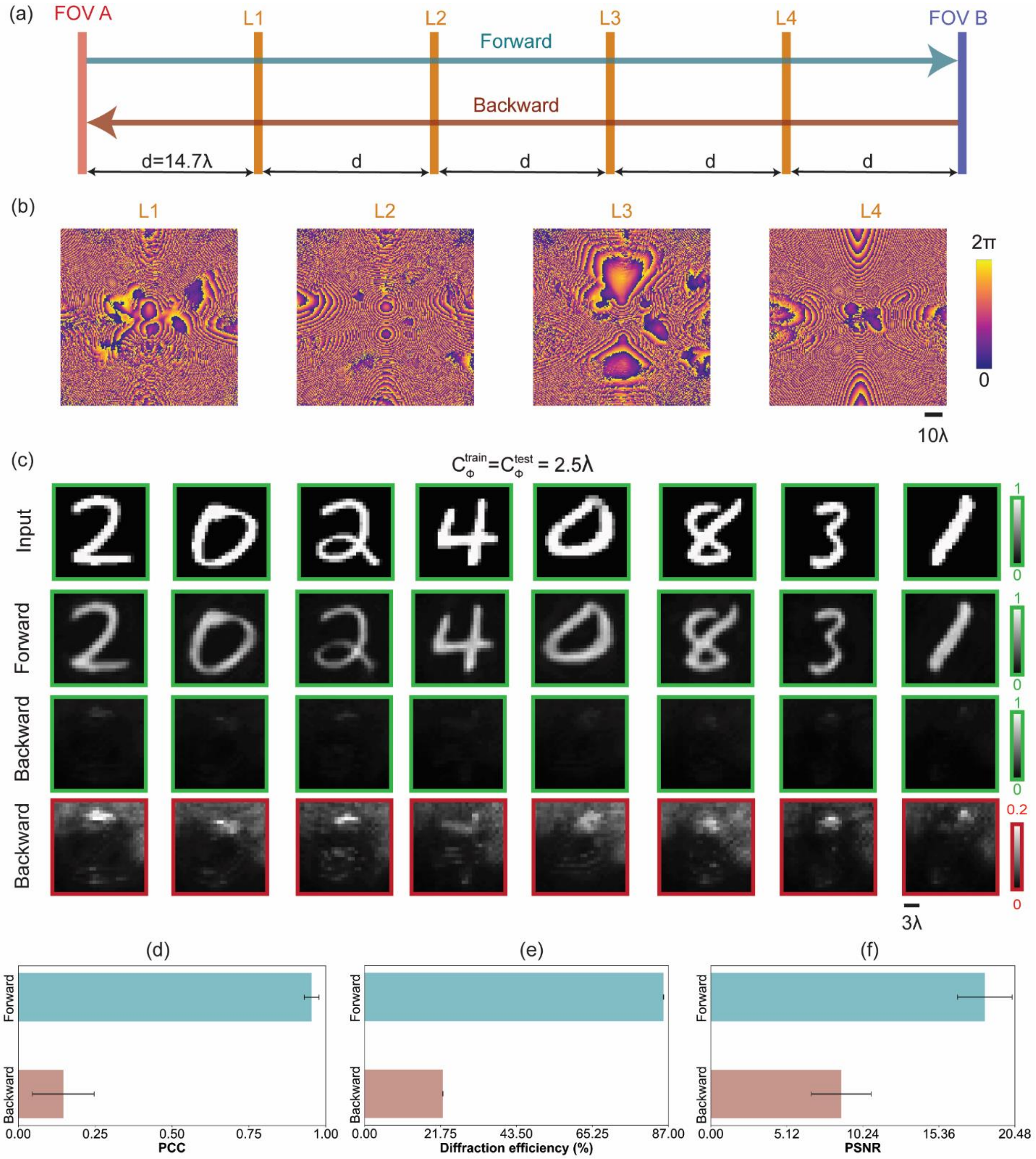

**Figure 2. Performance of a partially coherent unidirectional imager.** (a) Physical layout of a four-layer unidirectional imager design. (b) Optimized phase profiles of the diffractive layers in a unidirectional imager trained with $N_\Phi^{train} = 16$ and $C_\Phi^{train} = 2.5\lambda$. (c) Blind testing results of the unidirectional imager with $C_\Phi^{train} = C_\Phi^{test} = 2.5\lambda$ and $N_\Phi^{test} = 2048$. The first three rows display the input amplitude objects, forward output images, and backward output images, all using the same intensity range. For comparison,

the last row displays the backward output images with increased contrast. Note that both the forward propagation and backward propagation use the same input amplitude objects as displayed in the first row. (d)-(f) Performance evaluation of the diffractive unidirectional imager using 10,000 MNIST test images with PCC, diffraction efficiency, and PSNR metrics.

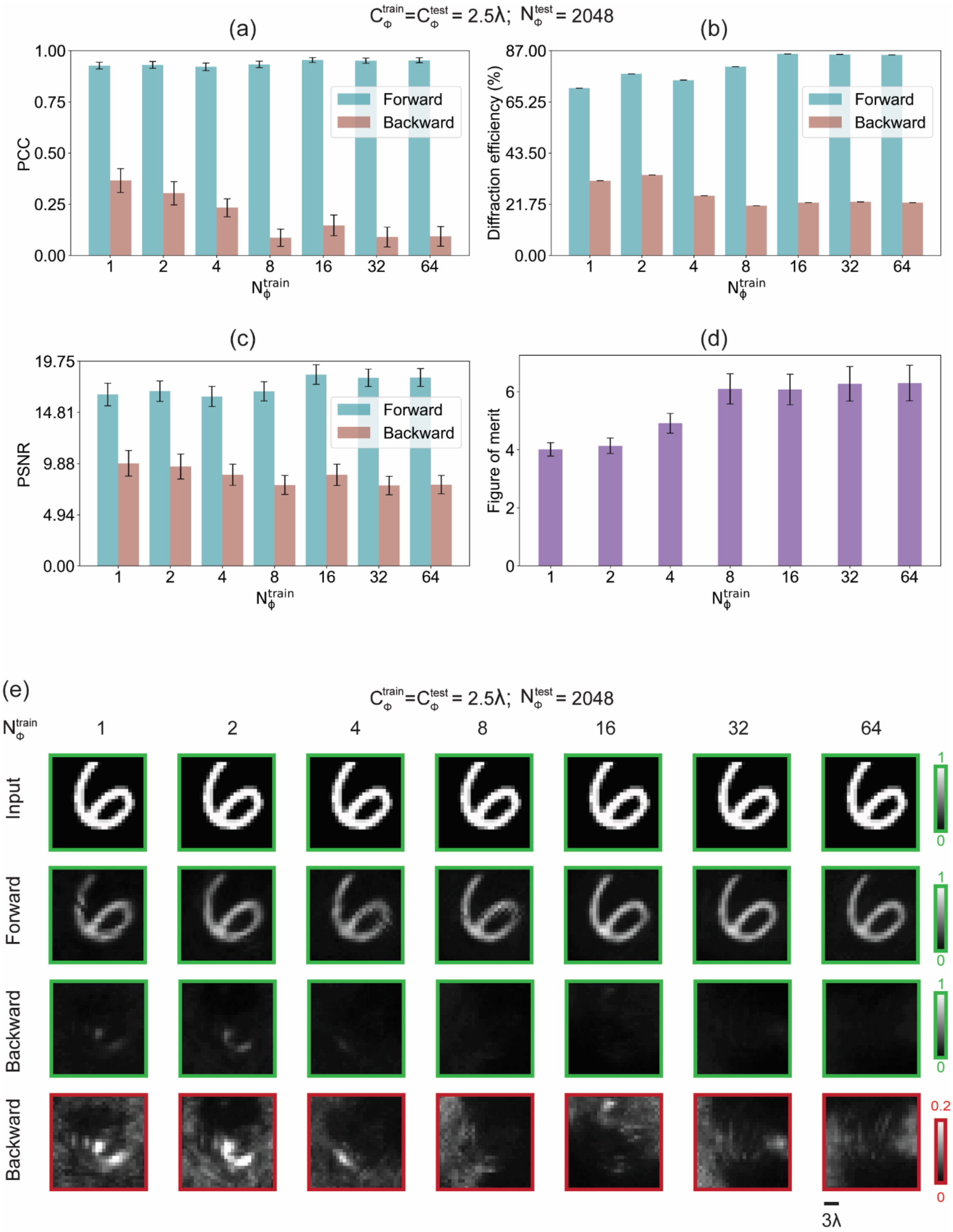

**Figure 3. Impact of $N_{\Phi}^{train}$ on the performance of partially coherent unidirectional imagers.** (a)-(d)

Performance analysis of partially coherent unidirectional diffractive imagers with $C_{\phi}^{train} = C_{\phi}^{test} = 2.5\lambda$, $N_{\phi}^{test} = 2048$, and different $N_{\phi}^{train}$ values ranging from 1 to 64. The performance in each case was evaluated using 10,000 MNIST test images with PCC, diffraction efficiency, PSNR, and FOM metrics. (e) Examples of the blinding testing results with different $N_{\phi}^{train}$ values. The first three rows display the input amplitude objects, forward output images, and backward output images, all using the same intensity range. For comparison, the last row shows the backward output images with increased contrast.

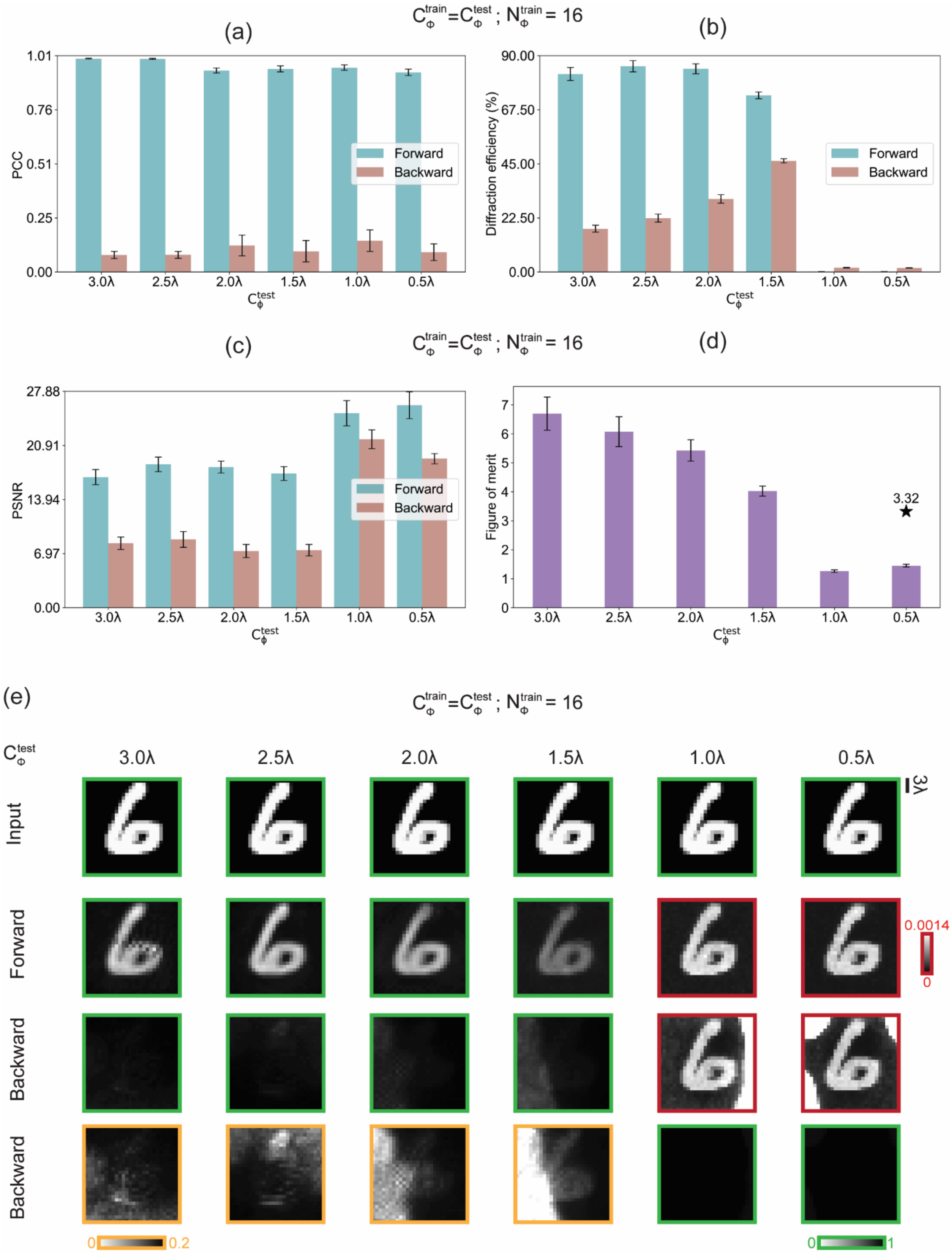

**Figure 4. Impact of $C_\Phi^{train}$ on the performance of partially coherent unidirectional imagers.** (a)-(d)

Performance analysis of partially coherent unidirectional diffractive imagers with different $C_{\phi}^{\mathrm{train}}$ values ranging from $\sim 3.0\lambda$ to $\sim 0.5\lambda$, where $C_{\phi}^{\mathrm{test}} = C_{\phi}^{\mathrm{train}}$. The performance in each case was evaluated using 10,000 MNIST test images with PCC, diffraction efficiency, PSNR, and FOM metrics. The star marker in (d) represents the diffractive model with $C_{\phi}^{\mathrm{train}} = 0.5\lambda$, which was trained using a high resolution image dataset composed of random intensity patterns. (e) Examples of the blinding testing results with different $C_{\phi}^{\mathrm{train}}$ values ranging from $\sim 3.0\lambda$ to $\sim 0.5\lambda$. The first three rows display the input amplitude objects, forward output images, and backward output images. For comparison, the last row shows the backward output images with different intensity ranges. Images with the same colored frame share the same intensity range.

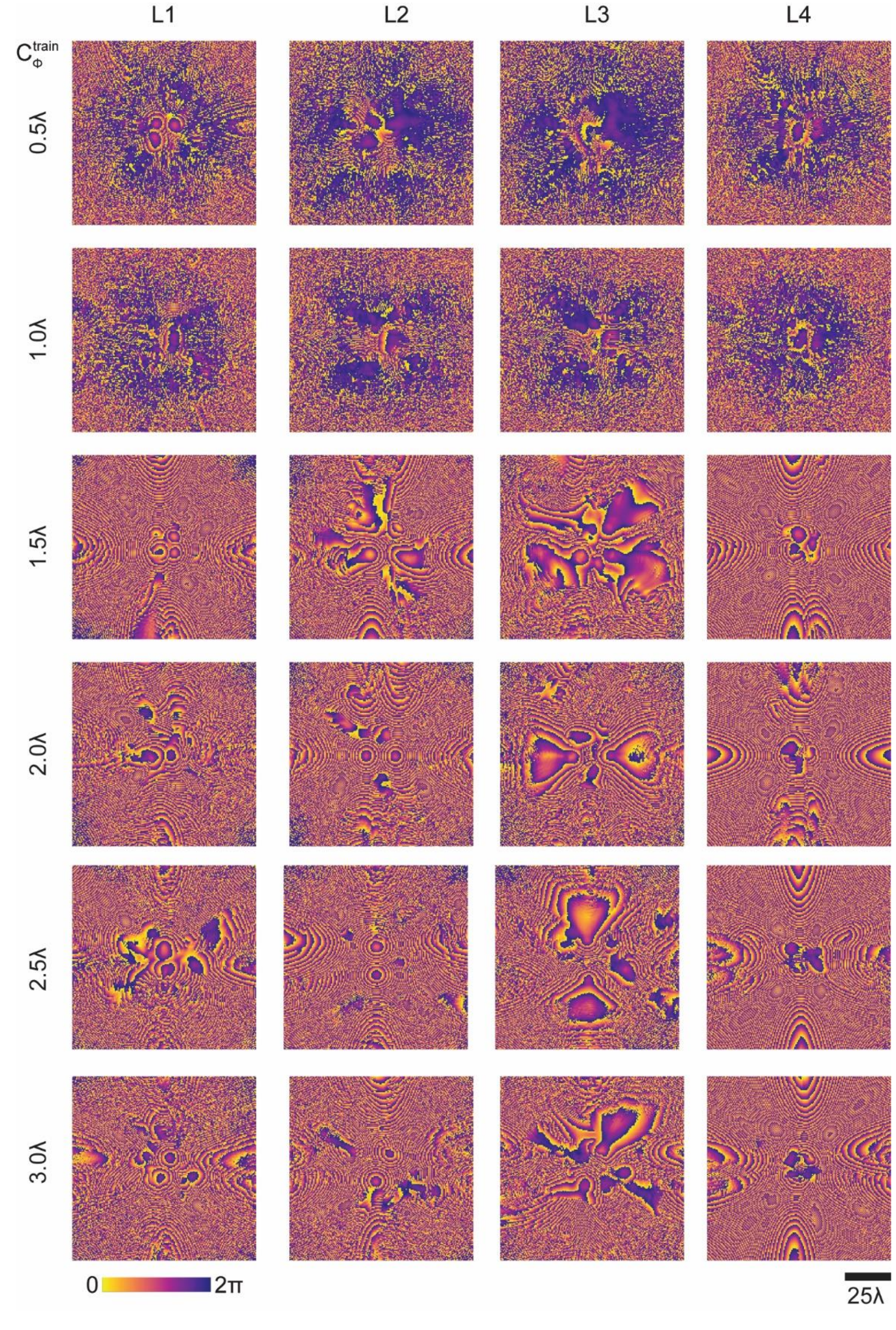

**Figure 5. Optimized phase profiles of the diffractive layers of different unidirectional imagers trained with varying $\mathbf{C_\phi^{train}}$ - from $\sim\mathbf{0.5\lambda}$ to $\sim\mathbf{3.0\lambda}$.**

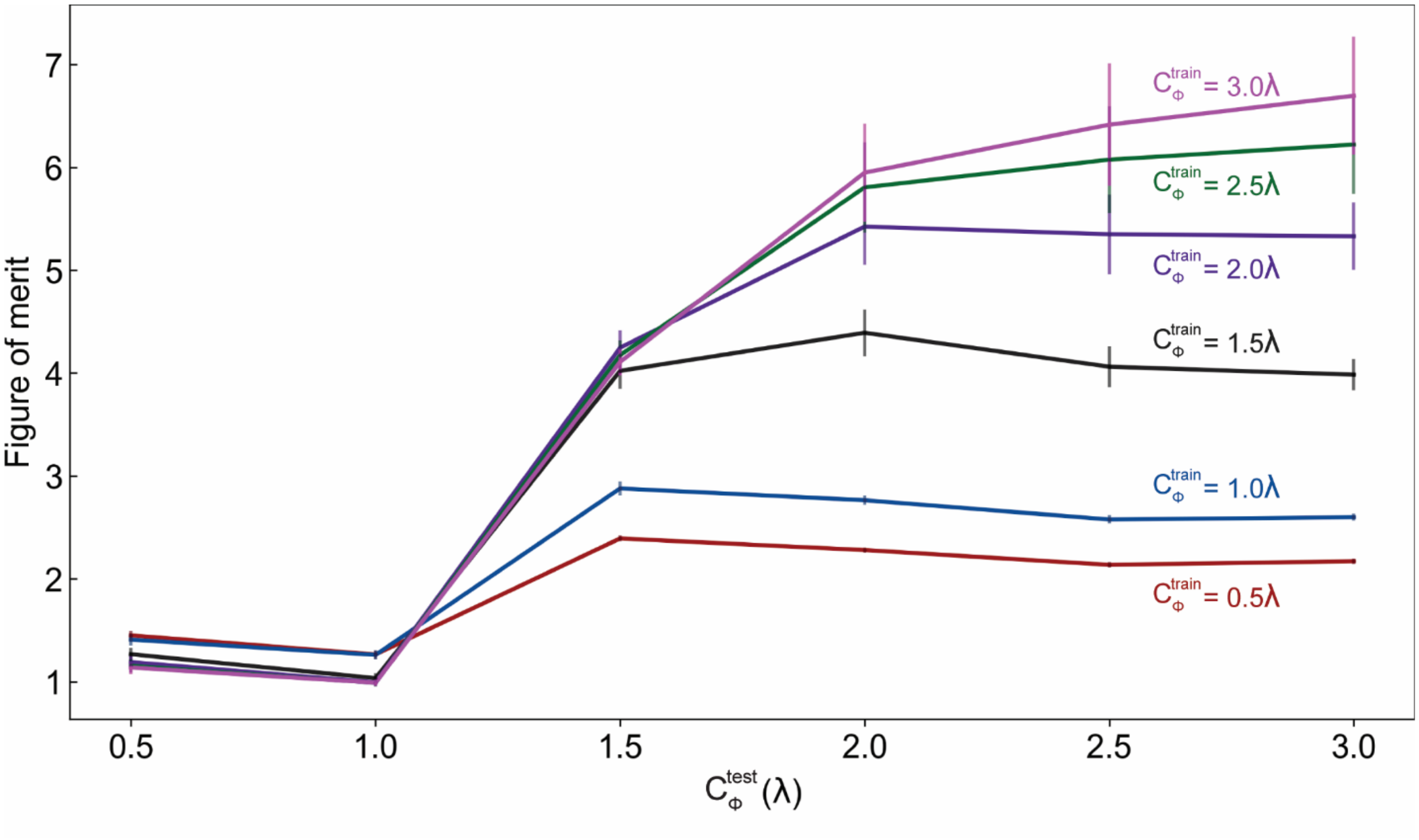

**Figure 6. The generalization performance of unidirectional diffractive imagers across various $\mathbf{C_\Phi^{test}}$ values, ranging from ~0.5$\lambda$ to ~3$\lambda$, despite being trained with a specific $\mathbf{C_\Phi^{train}}$.**

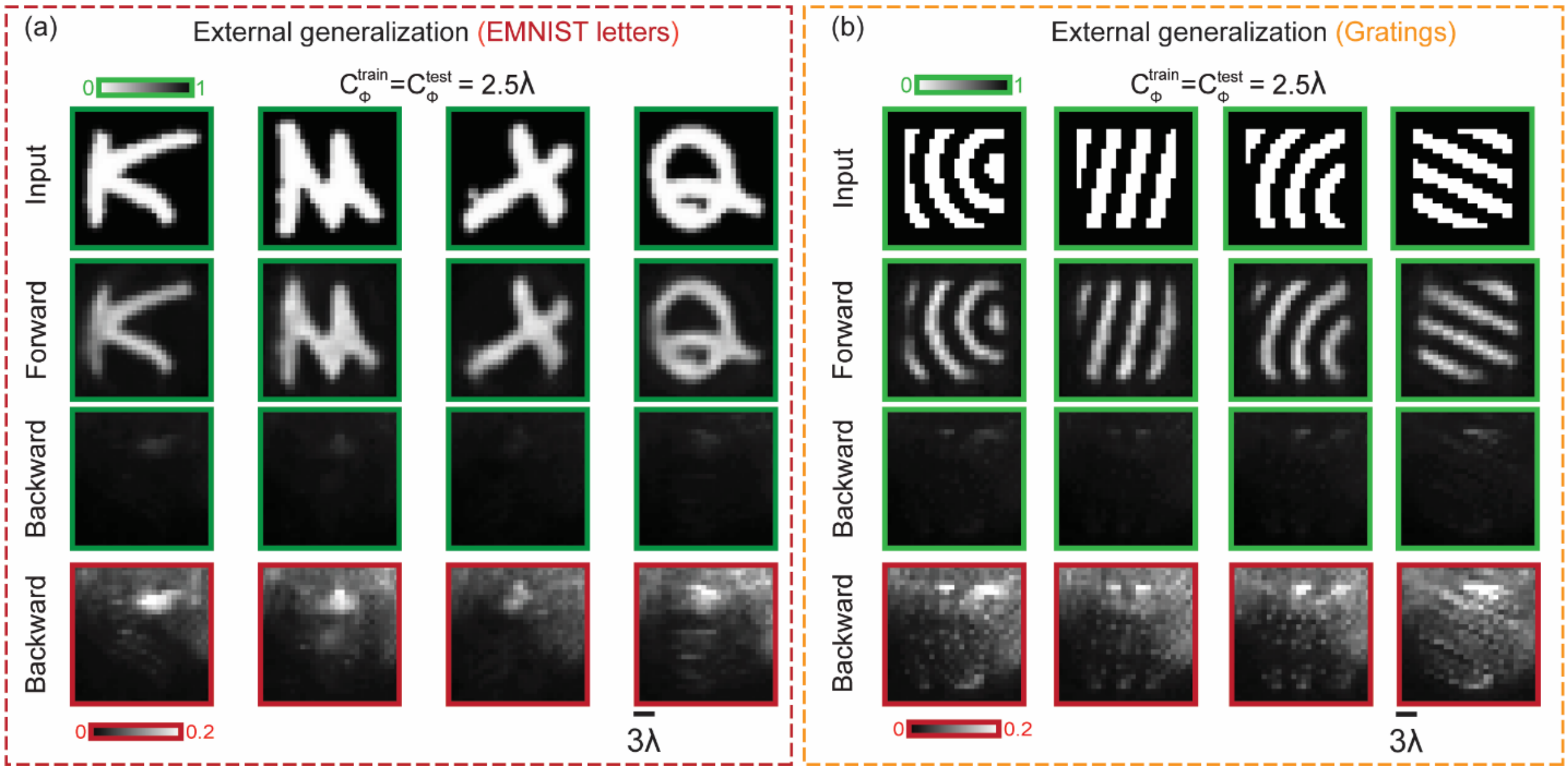

**Figure 7. Image dataset external generalization for the unidirectional imager with $\mathbf{C_\Phi^{train} = C_\Phi^{test} = 2.5\lambda}$.** Blinding testing results with EMNIST letters (a), and customized gratings (b), both of which were never used during training.

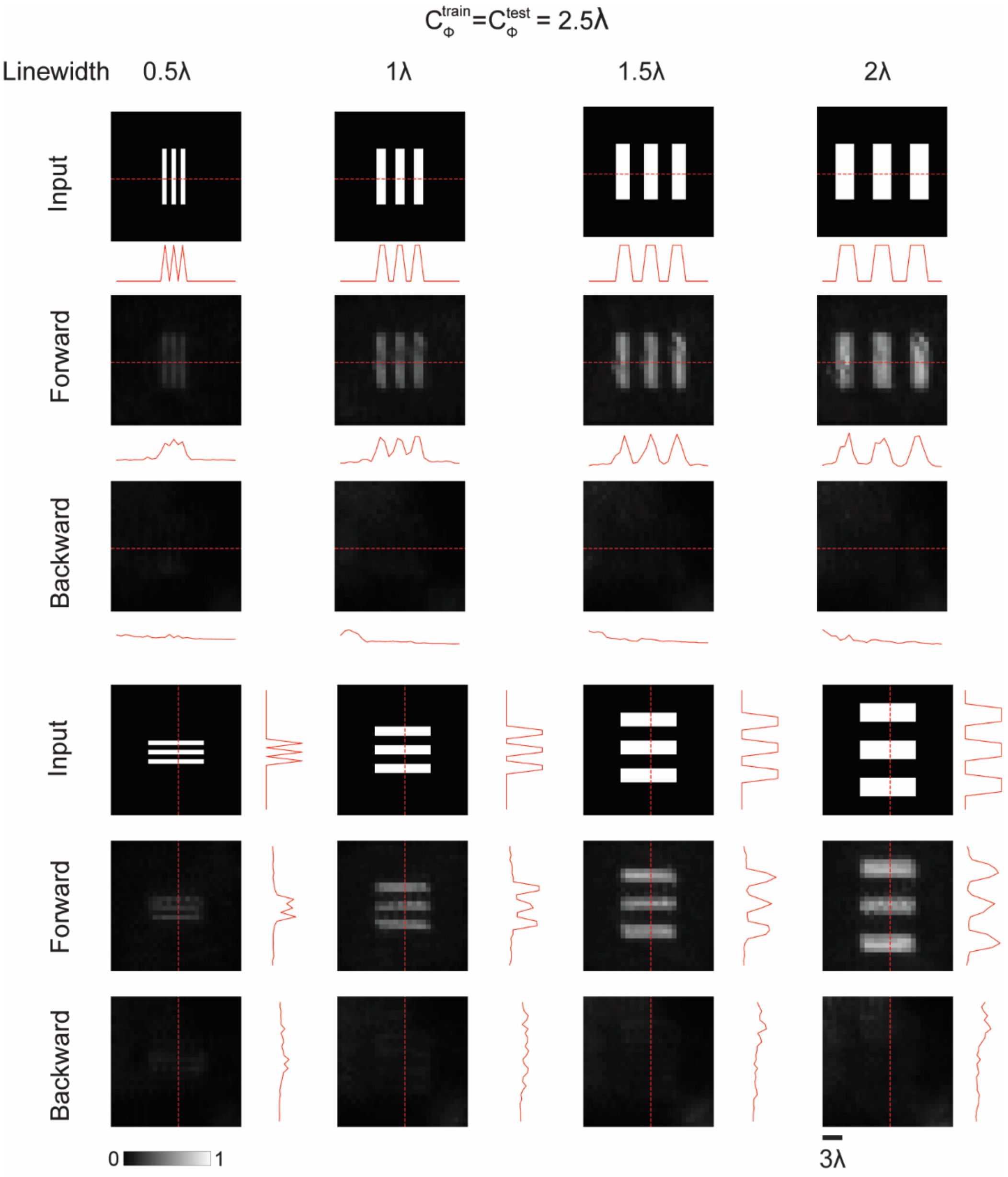

**Figure 8. Spatial resolution analysis for a unidirectional diffractive imager design with $C_{\Phi}^{train} = C_{\Phi}^{test} = 2.5\lambda$.**